\documentclass[12pt]{iopart}

\eqnobysec
\usepackage{epsf}
\usepackage{graphics}

\begin{document}

\title{Entanglement Witness Derived From NMR Superdense Coding}

\author{Robabeh Rahimi\dag, Kazuyuki Takeda\dag, Masanao Ozawa\ddag\ and Masahiro Kitagawa\dag
}

\address{\dag\ Graduate School of Engineering Science, Osaka University, 1-3 Machikaneyama, Toyonaka, Osaka 560-8531, JAPAN}

\address{\ddag\ Graduate School of Information Sciences, Tohoku University,\\ Aoba-ku, Sendai 980-8579, JAPAN}

\ead{rahimi@qc.ee.es.osaka-u.ac.jp}

\begin{abstract}
We show that it is possible to transfer two-bit information via encoding a single qubit in a conventional nuclear magnetic resonance (NMR) experiment with two very weakly polarized nuclear spins. Nevertheless, the experiment can not be regarded as a demonstration of superdense coding by means of NMR because it is based on the large number of molecules being involved in the ensemble state rather than the entanglement of the NMR states. Following the discussions, an entanglement witness, particularly applicable for NMR, is introduced based on separate and simultaneous measurement of the individual nuclear spin magnetizations.
\end{abstract}

\submitto{\JPA}

\maketitle

\section{Introduction}
A quantum computer, which is promised to be more powerful than the classical computers in solving some intractable problems by employing quantum algorithms, is not completely available yet, even though there are several proposals based on different quantum systems. Nuclear magnetic resonance (NMR) \cite{1} has been known as one of the most viable methods for demonstrating even relatively complicated quantum algorithms\cite{2}. However, the status of NMR has not been completely approved compared to the other candidates such as photons and ion traps, in that it treats an ensemble system composed of a large number of molecules. Moreover, the states are mixed rather than quantum pure states. Accordingly, nearly all previous attempts to demonstrate NMR quantum information processing (QIP) and quantum computations (QC) relied on pseudo-initialization\cite{3, 4} of highly mixed states. However, with or without employing the pseudo-pure states, mathematical arguments refute the credibility of NMR QIP and QC except at very low spin temperatures by verifying separability of the states\cite{5, 6}.

Entanglement is believed to be an essential requirement for quantum non-local algorithms to give enhancement over the classical counterparts\cite{7, 8}.  For example, superdense coding (SDC)\cite{9} enables transfer of two bits of classical information by encoding a single qubit. This remarkable effect is ascribed to the existence of entanglement in the system. However, in the implementation of the non-local algorithms by means of NMR, there is confusion with regard to the role of entanglement. NMR SDC has been demonstrated\cite{10}, as well as other quantum non-local algorithms\cite{2, 11}, with the separable states with which the non-local algorithms should not work.

In this work, we show that with NMR mixed states, even if the states are definitely separable, still there is a considerably high probability to detect signals with appropriate signs which apparently imply transfer of two-bit message even though we encode only one qubit.  The reason for detection of the signals is proved to be because of the large number of molecules being involved in the NMR ensemble state. Therefore, a system composed of a huge number of molecules should be necessarily prepared in order to realize the transfer of two bits of information. Then, taking account the required number of resources, molecules, it is totally non-sense to claim on the quantum advantage over the classical counterpart. However, for an exact demonstration of SDC with NMR, in the sense that it can realize enhancement over the classical communication, nuclear spin polarization should be increased above a certain threshold, and this threshold coincides with the mathematical criterion for non-separability of the density matrix.

We also discuss experimental detection of entanglement in terms of the concept of entanglement witness\cite{12, 13}. We introduce a new class of entanglement witness, which is based on measurement of nuclear spin magnetizations in a single run experiment. This approach provides a simple and convenient way of evaluating the existence of entanglement, and is applicable to all possible states encountered in SDC. Although the entanglement witness derived from the conventional approach is also shown to be measurable in a single run experiment, it requires pre-application of somewhat complicated unitary transformation that depends on the state under investigation.
\section{Ideal SDC with Nuclear Spins}
Let us consider a pair of nuclear spins $I=1/2$ and $S=1/2$ placed in a static magnetic field $B_0$, and suppose for a moment that the system is initially in a pure state $|\psi_0\rangle = | 00 \rangle$\cite{1}. Then, the procedure of SDC, whose quantum circuit is described in Fig.1, is as follows\cite{9}. 
\begin{figure}
\begin{center}
\scalebox{0.6}
{\includegraphics[0cm,21cm][20cm,27cm]{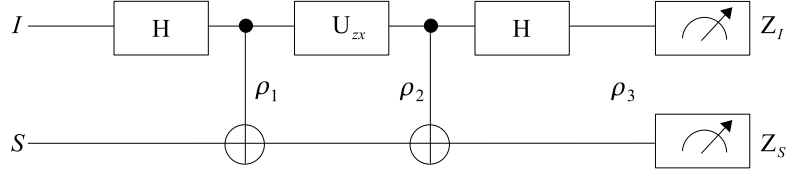}}
\end{center}
\caption{\label{Figure. 1} A quantum circuit for superdense coding. Two nuclear spins $I$ and $S$ are involved. After the first Hadamard and controlled-NOT, the spin $I$ is given to Bob and $S$ to Alice. Bob encodes the spin $I$ by applying the unitary transformation ${\rm U}_{zx}$ and then sends it off to Alice. Alice has now the two nuclear spins in her disposal. She applies the second controlled-NOT and Hadamard gates and measures the spin magnetizations. As a result, she obtains the encoded message, meaning that two-bit message is transferred while only a single spin has been encoded.}
\end{figure}
Firstly, the entangling operation is applied on the state of the nuclear spins. The entangling operation is composed of a Hadamard gate (H) on the $I$ spin and then a controlled-NOT gate (${\rm U_{cn}}$) whose control and target qubits are the $I$ and $S$ spins, respectively. The quantum state $ |\psi_1 \rangle $ after the entangling operation is represented as
\begin{equation}
\label{1}
  |\psi_1\rangle = {\rm U}_{\rm ent} |\psi_0\rangle =\frac{1}{\sqrt{2}} (|00\rangle + |11\rangle)= |\beta_{00}\rangle,
\end{equation}
where we have defined ${\rm U}_{\rm ent}={\rm U_{cn}} ({\rm H}_I\otimes {\rm I}_S)$. Here, $ | \beta_{00}\rangle$ is known as one of the four Bell states\cite{12}
\begin{equation}
\label{2}
|\beta _{zx}\rangle\equiv\frac{|0,x\rangle+(-1)^z|1,\bar{x}\rangle}{\sqrt 2},
\end{equation}
where $z, x=0, 1$ and $\bar{x} = 1-x$. Now, suppose that the nuclear spin $I$ is given to Bob and the other nuclear spin to Alice. Bob encodes a two-bit classical message $zx$ by applying the unitary operation ${\rm U}_{zx}={\rm Z}^z {\rm X}^x$ on the $I$ spin and then sends off the encoded qubit to Alice. The effect of the unitary transformation by Bob is to toggle $|\psi_1\rangle = |\beta_{00}\rangle$ into the other Bell state. That is,
\begin{eqnarray}
\label{3}
  |\psi_2\rangle = {\rm U}_{zx} |\beta_{00}\rangle
                 = |\beta_{zx}\rangle.
\end{eqnarray}
Then, Alice applies the disentangling operation, which is composed of a ${\rm U_{cn}}$ followed by ${\rm H}$. The state $|\psi_3\rangle$ after the disentangling operation is
\begin{equation}
\label{4}
  |\psi_3\rangle = {\rm U}_{\rm Bell}|\psi_2\rangle=|zx\rangle,
\end{equation}
where we defined ${\rm U}_{\rm Bell}=({\rm H}_I\otimes {\rm I}_S) {\rm U_{cn}}$. Finally, she performs measurement of the resultant magnetizations, ${\rm Z}_I$ and ${\rm Z}_S$ and extracts the results as $(-1)^z$ and $(-1)^x$.

If it was possible to execute all the above process in an NMR apparatus, then the signal intensities detected through the measurement in this ideal case of the fully polarized initial state would have been maximum and information on the choice of $ {\rm U}_{zx} $ could be extracted with a success probability of unity.   In this case,  the maximum entanglement is shared between Alice and Bob, which causes transfer of two classical bits of information from Bob to Alice via the $I$ spin alone. Communication by SDC is efficient by a factor of two as compared to the classical communication.

\section{Transfer of Two-Bit Message by Encoding a Single Spin in NMR}
NMR \cite{1} is a spectroscopy, manipulating the nuclear spins placed in a strong, typically 10 Tesla, static magnetic field $B_0$. The corresponding total spin Hamiltonian in an isotropic liquid state is composed of several terms, but the most important parts are considered as the Zeeman Hamiltonian and spin-spin interaction parts. So that we take
\begin{equation}
H=- \sum_{i=0}^{m-1} \frac{\hbar \omega_{i}}{2} {\rm Z}_i +\frac{1}{4} \sum_{i=0}^{m-2} \sum_{j>i} 2 \pi \hbar J_{ij} {\rm Z}_i {\rm Z}_j,
\end{equation}
where $\omega_{i}$ is the Larmor frequency and ${\rm Z}_i$ is the Pauli operator for the ${\rm z}$ component of the $i$th spin and $J_{ij}$ is the coupling constant between $i$th and $j$th spins in a molecule. In the high field approximation, which is valid for most practical cases, in the thermal equilibrium at temperature $T$, the polarization $\epsilon_i$ is given by
\begin{equation}
\label{epsilon}
\epsilon_i={\rm tanh} ( \frac{-\gamma_i\hbar B_0}{2k_{\rm B}T} ),
\end{equation}
where $\gamma_i$ is gyromagnetic ratio. Accordingly, the probabilities $p_i$ and $q_i$ of finding the $i$th spin in the states $|0\rangle$ and $|1\rangle$, respectively, are given by
\begin{equation}
\label{probability}
p_i=\frac{1+\epsilon_i}{2},\, \hspace{1cm} q_i=\frac{1-\epsilon_i}{2}.
\end{equation}
Then, the density matrix in thermal equilibrium is given by
\begin{equation}
\label{ron}
\rho_{\rm m-qubit}= \rho^1 \otimes ... \otimes \rho^m=\otimes_{i=1}^m\rho^{i},
\end{equation}
where $\rho^{i}$ is the density matrix for the $i$th nuclear spin in a molecule and is written as
\begin{equation}
\label{roi}
\rho^{i} =  p_i | 0 \rangle \langle 0 | + q_i | 1 \rangle \langle 1 |.
\end{equation}
At room temperature, the energy splitting between the ground state and the excited state is much smaller than the thermal energy and as a result, large number of spins (qubits) are in statistical mixture with low spin polarization. Accordingly, the information processing should be evaluated statistically.

For nuclear spins $I=1/2$ and $S=1/2$ the initial state is a mixed state described by partially polarized density matrix with the form of (\ref{ron}) and (\ref{roi}) with ${\rm m=2}$ as
\begin{eqnarray}
\label{roin}
\rho_{0}&=&(p_I|0\rangle\langle 0|+q_I|1\rangle\langle 1|)\otimes (p_S|0\rangle\langle 0|+q_S|1\rangle\langle 1|)\\ \nonumber
        &=&p_Ip_S|00\rangle\langle00|+p_Iq_S|01\rangle\langle01|+q_Ip_S|10\rangle\langle10|+q_Iq_S|11\rangle\langle11|.
\end{eqnarray}
After performing the entangling operation the state of the two qubits (\ref{roin}) changes to a general Bell diagonal state of the form
\begin{eqnarray}
\label{roshared}
\rho_{1}&=& {\rm U}_{\rm ent} \rho_0 {\rm U}^\dagger_{\rm ent}\\ \nonumber
        &=& p_Ip_S |\beta _{00}\rangle \langle \beta_{00}|+p_I q_S|\beta _{01}\rangle \langle \beta_{01}|+\\ \nonumber 
        & & q_Ip_S|\beta_{10}\rangle \langle \beta_{10}|+q_Iq_S|\beta _{11}\rangle \langle \beta_{11}|.
\end{eqnarray}
This is the shared state between Alice and Bob. Suppose that Bob owns the state of the nuclear spin $I$. Then, Bob applies the unitary operation ${\rm U}_{zx}$. The state $\rho_2$, after the encoding operation is still a general Bell diagonal state as follows
\begin{eqnarray}
\label{encoded}
\rho_{2}&=& {\rm U}_{zx} \rho_1 {\rm U}^\dagger_{zx}\\ \nonumber
        &=& p_Ip_S |\beta _{z, x}\rangle \langle \beta_{z, x}|+p_Iq_S|\beta _{z, {\bar x}}\rangle \langle \beta_{z, {\bar x}}|+\\ \nonumber
        & & q_Ip_S|\beta _{{\bar z}, x}\rangle \langle \beta_{{\bar z}, x}|+q_Iq_S|\beta _{{\bar z}, {\bar x}}\rangle \langle \beta_{{\bar z}, {\bar x}}|.
\end{eqnarray}
He hands over the nuclear spin to Alice, who applies the disentangling operation.  Then, the state $\rho_2$ changes into
\begin{eqnarray}
\rho_{3}&=& {\rm U}_{\rm Bell} \rho_2 {\rm U}^\dagger_{\rm Bell}\\ \nonumber
        &=& p_Ip_S |z, x\rangle \langle z, x|+p_Iq_S|z, {\bar x}\rangle \langle z, {\bar x}|+\\ \nonumber
        & & q_Ip_S|{\bar z}, x\rangle \langle {\bar z}, x|+q_Iq_S|{\bar z}, {\bar x}\rangle \langle {\bar z}, {\bar x}|\\ \nonumber
        &=& (p_I|z\rangle \langle z|+q_I|{\bar z}\rangle \langle {\bar z}|)\otimes (p_S|x\rangle \langle x|+q_S|{\bar x}\rangle \langle {\bar x}|).
\end{eqnarray}
Measurement of the spin magnetizations $ {\rm Z}_I $ and $ {\rm Z}_S $ is done on the total ensemble state composed of $n$ molecules. Then, measurement on the product state $ \otimes_{i=0}^n \rho^{(i)}_3$ is composed of separate but simultaneous measurements, $\sum_{i=1}^n {\rm Z}_I^{(i)}$ and $\sum_{i=1}^n {\rm Z}_S^{(i)}$, on the spin magnetizations. Note, that $\rho^{(i)}_3$ stands for the density matrix $\rho_3$ of the $i$th molecule. Measurement of the spin magnetizations gives results as binomial probability distributions over $(-n, -n+1, ..., -1, 0, 1, ..., n-1, n)$, with the mean values to be as follows
\begin{eqnarray}
\label{mean}
\mu_I&=&(-1)^z n p_I+(-1)^{\bar z} n q_I=(-1)^z n\epsilon_I,\\ \nonumber
\mu_S&=&(-1)^x n p_S+(-1)^{\bar x} n q_S=(-1)^x n\epsilon_S.
\end{eqnarray}
Let us make an assumption that $z, x=0$, which does not influence the generality of the discussion. The corresponding variances are characterized by
\begin{eqnarray}
\label{var}
\sigma_I^2&=&4np_Iq_I=n(1-\epsilon_I^2),\\ \nonumber
\sigma_S^2&=&4np_Sq_S=n(1-\epsilon_S^2).
\end{eqnarray}
Recall that $\sigma_I$ and $\sigma_S$ are essentially the widths of the range over which the outcomes are distributed around the mean values (\ref{mean}). The relative widths of distributions are characterized by
\begin{eqnarray}
\frac{\sigma_I}{\mu_I}=\frac{\sqrt{n(1-\epsilon_I^2)}}{n\epsilon_I}\approx \frac{1}{\epsilon_I\sqrt n},\\ \nonumber
\frac{\sigma_S}{\mu_S}=\frac{\sqrt{n(1-\epsilon_S^2)}}{n\epsilon_S}\approx \frac{1}{\epsilon_S\sqrt n}.
\end{eqnarray}
Then it is clear that the relative widths decrease as $\sqrt n$ with increasing  the number of molecules $n$. Thus, the greater the number of molecules, the more likely it is that an observation gives a result which is relatively close to the mean values (\ref{mean}).

Now, it is also required to calculate the corresponding error probability of detection of a negative value for $z=0$. For simplicity, let us consider only ${\rm Z}_I^{(i)}$, which is the measurement on the spin magnetization of the nuclear spin $I$ in the $i$th molecule. It should be clear at the moment that results are applicable to the signals detected through measurement on the other spin magnetization in the molecule, ${\rm Z}_S^{(i)}$. The error probability is defined as
\begin{equation}
P_e=P(\sum_{i=1}^n{\rm Z}^{(i)}_I<0|z=0).
\end{equation}
We would like to calculate the error probability to show that for a range of $n$ of the current NMR experiments, this quantity is negligible. From the DeMoivre and Laplace theorem\cite{15} if $n$ is large enough then generally we have
\begin{equation}
P\left\{\alpha<\frac{\sum_i^n{\rm Z}^{(i)}_I-{\mu_I}}{{\sigma_I}}<\beta\right\}\approx\frac{1}{\sqrt{2\pi}}\int_{\alpha}^{\beta}e^{-\frac{x2}{2}}dx.
\end{equation}
For the error probability $P_e$, therefore
\begin{eqnarray}
P_e&=&P\left\{-\infty<\frac{\sum_i^n{\rm Z}^{(i)}_I-{\mu}_I}{\sigma_I}<-\frac{\mu_I}{\sigma_I}\right\}\\ \nonumber
   &\approx&\frac{1}{\sqrt{2\pi}}\int_{-\infty}^{-\mu_I/\sigma_I}e^{-\frac{x2}{2}}dx\\ \nonumber
   &\approx&\frac{1}{\sqrt{2\pi}}e^{-\frac{{\mu_I}^{2}}{2{\sigma_I}^2}}\frac{{\sigma_I}}{\mu_I}\\ \nonumber
   &\approx&\frac{1}{\sqrt{2\pi}}e^{-\frac{({n\epsilon_I^2})}{2}}\frac{1}{({\sqrt n \epsilon_I})}.
\end{eqnarray}
If the number of molecules are large enough, the result of the measurement on the spin magnetizations $ {\rm Z}_I $ and $ {\rm Z}_S $ on the ensemble state gives results very close to the mean values (\ref{mean}) with negligible error probabilities. In a conventional NMR experiment, the number of molecules are $n\sim 10^{18}$, with low spin polarization $\epsilon\sim 10^{-5}$. Then we calculate $P_e\ll 10^{-100}$, to be negligible. Therefore, in an NMR apparatus, if signals can be detected despite of the low spin polarization the choice $zx$ can be evaluated with probability of almost one through the signs of the signals (\ref{mean}). 

Detection of signals depends on the NMR apparatus. The NMR signal intensity is defined by amplitude
\begin{equation}
\label{S}
V_{\rm S}=\frac{1}{4}\sqrt{(Q/V)\mu_0R\omega_I}\hbar \gamma_In\epsilon_I,
\end{equation}
and the noise is determined through Nyquist formula\cite{16}
\begin{equation}
\label{N}
V_{\rm N}=\sqrt{4k_{\rm B}TR\Delta\nu},
\end{equation}
where $Q$ is the quality factor of the resonance coil, $V$ is the volume of the coil, $R$ is the resistance, $\omega_I$ is the Larmor frequency of the nuclear spin $I$, and $\Delta\nu$ is the amplifier bandwidth. The noise which is defined by (\ref{N}) is entirely classical noise generated by the equilibrium fluctuations of the electric current inside an electrical conductor. Nevertheless, for an NMR experiment with two qubit liquid ensemble at room temperature with $V=1{\rm cm}^3$ and $Q=10^3$, the number of required molecules is bounded by $N>10^{16}$\cite{17}.

Therefore, in the case which we are facing, $n$ is large to get strong enough signal intensities to be detectable with an NMR apparatus. Deviations are very small (\ref{var}) and the signals are very close to the mean values (\ref{mean}) with the error probabilities to be negligible. Thus the signs of the detected signals give information on the choice of $zx$, and hence the encoded two-bit message, regardless of the separability due to the very low polarized initial states. Then we emphasize the results of the measurement of the spin magnetizations ${\rm Z}_I$ and ${\rm Z}_S$ to be as follows
\begin{equation}
\label{result}
\langle {\rm Z}_I \rangle=(-1)^{z}\epsilon_I,\hspace{1.5cm} \langle{\rm Z}_S \rangle=(-1)^{x}\epsilon_S.
\end{equation}
We note that this does not necessarily mean that the experiment described here is a demonstration of SDC with NMR mixed states, even though the signs of the detected signals appear to give the two-bit message while the $I$ spin alone has been applied by encoding pulses. SDC relies on entanglement of states in order to realize transfer of two-bit information by only encoding a single qubit. In the case of the NMR experiment, as explained here, the large number of molecules are inevitably required for detection of encoded two-bit message. Then taking account the required number of molecules it is in principle erroneous to call the corresponding NMR experiment a demonstration of SDC.
\section{Detection of Entanglement}
Given the density matrix (\ref{encoded}), we obtain a probability for a successful SDC to be $p_Ip_S$, with which a two-bit message can be transferred inside any individual single molecule. Therefore, for this experiment to represent the quantum information advantage and to outperform the classical one-bit communication through one-qubit channel, the success probability, $p_Ip_S$, must exceed $50\%$. This is because two-bit information can be obtained by one-bit classical communication and a one-bit of random guess, which comes true with a probability of $1/2$. Then SDC is beyond the classical achievements only if the inequality
\begin{equation}
\label{16}
p_Ip_S>1/2
\end{equation}
is satisfied. It is worth noting that this condition exactly coincides with the condition for the non-separability of $\rho_2$ derived from e.g. the negativity criterion\cite{18, 19}.

Therefore, only when $F$, to be defined as follows, imposes a negative value, the NMR SDC is successful and there exists entanglement.
\begin{equation}
\label{17}
F \equiv 1/2 - p_I p_S.
\end{equation}
Detection of entanglement through finding a negative value for an observable is reminiscent of entanglement witness\cite{12, 13}. Entanglement witness is a Hermitian operator $W=W^{\dagger}$ which has positive mean values for all separable states $\rho$, $\mathrm{Tr}(W\rho)>0$, but a negative mean value for at least one entangled state $\sigma_{\rm ent}$, $\mathrm{Tr}(W\sigma_{\rm ent})<0$. In other words, entanglement is detected if a negative mean value is obtained through the measurement of the entanglement witness.

Using the spin polarizations, we rewrite $F$ as
\begin{eqnarray}
\label{20}
F&=&\frac{1}{2} - \frac{1}{4} (1+\epsilon_I)(1+\epsilon_S)\\ \nonumber
 &=&\frac{1}{2} - \frac{1}{4} (1+|\langle {\rm Z}_I \rangle |)(1+|\langle {\rm Z}_S \rangle | ),
\end{eqnarray}
where we used (\ref{result}). The absolute values are required for the evaluation of the function $F$ for different choices of ${zx}$.

Measurement on the state $\rho_3$ with the observables ${\rm Z}_I$ and ${\rm Z}_S$ is equivalent to measurement on the state $\rho_2$ (or $\rho_1$ in the special case of $x, z=0$) in the Bell basis, because
\begin{eqnarray}
\label{23}
\langle {\rm Z}_I \rangle&=&\mathrm{Tr} \rho_3 ( {\rm Z}_I \otimes {\rm I}_S)\\ \nonumber
                         &=&\mathrm{Tr} \rho_2 ( {\rm X}_I \otimes {\rm X}_S)=\langle W_1 \rangle 
\end{eqnarray}
\begin{eqnarray}
\label{24}
\langle {\rm Z}_S \rangle&=&\mathrm{Tr} \rho_3 ( {\rm I}_I \otimes {\rm Z}_S)\\ \nonumber
                         &=&\mathrm{Tr} \rho_2 ( {\rm Z}_I \otimes {\rm Z}_S)=\langle W_2 \rangle ,
\end{eqnarray}
where the two observables $W_1$ and $W_2$ are defined as follows
\begin{equation}
\label{25}
W_1={\rm U}_{\rm Bell}^{\dagger} ( {\rm Z}_I \otimes {\rm I}_S) {\rm U}_{\rm Bell}={\rm X}_I \otimes {\rm X}_S, 
\end{equation}
\begin{equation}
\label{26}
W_2={\rm U}_{\rm Bell}^{\dagger} ( {\rm I}_I \otimes {\rm Z}_S) {\rm U}_{\rm Bell}={\rm Z}_I \otimes {\rm Z}_S.
\end{equation}
From (\ref{20}), (\ref{23}) and (\ref{24}), then $F$ is further rewritten as
\begin{equation}
\label{27}
F \equiv f(\langle W_1 \rangle,\langle W_2 \rangle)=\frac{1}{2} - \frac{1}{4} (1+|\langle W_1 \rangle |)(1+|\langle W_2 \rangle | ).
\end{equation}
Measurement of the two observables $W_1$ and $W_2$ on the ensemble system described by $\rho_2$ is related to the measurement on the state $\rho_3$ of the spin magnetization, which affords information on the spin polarizations. Therefore, separate and simultaneous measurement of the observables $W_1$ and $W_2$ is possible in a single experiment, and tells the existence of entanglement. That is, if $\langle W_1 \rangle$ and $\langle W_2 \rangle$ satisfy
\begin{equation}
\label{28}
F \equiv f(\langle W_1 \rangle,\langle W_2 \rangle)<0,
\end{equation}
then the state is entangled. In this sense, $W_1$ and $W_2$ are regarded as a new class of entanglement witnesses. These observables are easily measurable in a single run NMR experiment and give information on the status of entanglement through quantitative evaluation of the function $F$ (\ref{27}).

We also note, that entanglement can be detected in principle through measuring the conventional entanglement witness\cite{12, 13}.  For $\rho_2$, entanglement witness is derived by the conventional approach\cite{12, 13} as
\begin{equation}
\label{32}
W=\frac{1}{4} ({\rm I}_I \otimes {\rm I}_S+(-1)^{\bar z}{\rm X}_I \otimes {\rm X}_S +(-1)^{\bar z}(-1)^{\bar x}{\rm Y}_I \otimes {\rm Y}_S +(-1)^{\bar x}{\rm Z}_I \otimes {\rm Z}_S).
\end{equation}
If we assume the ability of implementation of any form of the unitary transformations with NMR pulse sequences, then the conventional entanglement witness also can be measured in a single run experiment and by measuring the spin magnetizations (see appendix A). We emphasize here, that the scheme which is introduced in this contribution for detection of entanglement, still has advantage in the sense that for different choices of $z$ and $x$ it is applicable by just evaluation the function with the absolute values. In other words, there is no need to change the experimental operations. However, the conventional entanglement witness even though is proved here to be measurable in a single run NMR experiment by only measuring the spin magnetizations, still requires different pre-applied unitary transformations, which depend on the choices of $z$ and $x$.

\section{Conclusion}
Liquid state NMR with very low nuclear spin polarizations prohibits the existence of entanglement. Then, with this physical system, it is absolutely impossible to demonstrate a faithful non-local quantum information processing, which requires entangled states. Although two-bit information is correctly detected in NMR SDC experiment, it is not based on the existence of entanglement but relys on the large number of molecules.

For a completely reliable demonstration of NMR SDC within a single molecule, spin polarization should inevitably be enhanced over a certain threshold and this threshold coincides with the condition for non-separability of the states. According to the results, we introduced a new class of entanglement witnesses, with several advantages particularly for NMR. The introduced entanglement witness is measurable in a single run experiment and generally is applicable for all the states without any requirement on any extra experimental operation. Detection of entanglement through the conventional entanglement witness is also proved to be possible in a single NMR experiment, however for different states under investigation, it requires different pre-application of somehow complicated unitary transformations.

\ack
R. R. and M. K. are grateful for helpful discussions with Dr. Fumiaki Morikoshi. R.R. would like to thank Akira SaiToh for contributions to the appendix A. This work has been supported by CREST of Japan Science and Technology Agency.
\appendix
\setcounter{section}{1}
\section*{Appendix A. Single Run Detection of the Conventional Entanglement Witness}
Suppose that we have the observable $\tilde W={\rm U}^\dagger \tilde W_o {\rm U}={\rm U}^\dagger(a{\rm Z}_I\otimes {\rm I}_S+b{\rm I}_I\otimes {\rm Z}_S+c{\rm I}_I\otimes {\rm I}_S){\rm U}$ with coefficients $a,b,c \in \mathbf{R}$ and a unitary transformation ${\rm U}\in \mathrm{U}(4)$, and have the conventional entanglement witness (\ref{32}). For $zx$, which is fixed to each of the possible four classes, we would like to prove with an explicite example that there exists a set of $a,b,c$, and ${\rm U}$ such that $\mathrm{Tr} \rho_2 \tilde W =\mathrm{Tr} \rho_2 W$.

In the following, we first restrict the problem to the case of $x,z=0$ then will discuss on the other cases. The entanglement witness $W$ for $z=x=0$ is\cite{12, 13}
\begin{equation}
\label{conen}
W=\frac{1}{4}({\rm I}_I\otimes {\rm I}_S -{\rm X}_I\otimes {\rm X}_S + {\rm Y}_I\otimes {\rm Y}_S - {\rm Z}_I\otimes {\rm Z}_S).
\end{equation}
The eigenvalues of $W$ are different to the ones for $\tilde W$. Then the two observables are essentially different to each other. However they can still give the same value after calculating the traces.

Consider the case where $p_I=p_S=1/2$, a maximally mixed state $\rho_2=\frac{1}{4}({\rm I}_I \otimes {\rm I}_S)$ for the input. We have
\begin{eqnarray}
 \mathrm{Tr}\rho_2W&=&\frac{1}{4},\\ \nonumber
 \mathrm{Tr}\rho_2\tilde W&=&\mathrm{Tr}{\rm U}\rho_2{\rm U}^\dagger\tilde W_o=c.
\end{eqnarray}
Therefore $c=1/4$ holds. An equivalent problem, then would be finding a set of $\alpha=a+b$ and $\beta=a-b$ that satisfies
\begin{equation}
 \mathrm{Tr}\rho_1(W'-{\rm V}^\dagger \tilde W_o{\rm V})=0,
\end{equation}
where ${\rm V}={\rm U}{\rm U_{cn}}({\rm H}_I\otimes{\rm I}_S)={\rm U}{\rm U}_{\rm ent}$ and $ W'={\rm U}_{\rm ent}^\dagger W {\rm U}_{\rm ent}$. Because $\rho_1$ is a diagonal matrix, the above equality holds for $\forall ~p_I,p_S$ if the diagonal elements of $(W'-{\rm V}^\dagger \tilde W_o{\rm V})$ are zero. This condition is expressed as follows
\begin{equation}
 W'-{\rm V}^\dagger \tilde W_o{\rm V}=
\left(
\begin{array}{@{\,}cccc@{\,}}
0&a_{01}&a_{02}&a_{03}\\
a_{01}^*&0&a_{12}&a_{13}\\
a_{02}^*&a_{12}^*&0&a_{23}\\
a_{03}^*&a_{13}^*&a_{23}^*&0
\end{array}
\right).
\end{equation}
This leads to that
\begin{equation}
\label{A}
A\equiv \left(
\begin{array}{@{\,}cccc@{\,}}
-\frac{3}{4}&-a_{01}&-a_{02}&-a_{03}\\
-a_{01}^*&\frac{1}{4}&-a_{12}&-a_{13}\\
-a_{02}^*&-a_{12}^*&\frac{1}{4}&-a_{23}\\
-a_{03}^*&-a_{13}^*&-a_{23}^*&\frac{1}{4}
\end{array}
\right)
={\rm V}^\dagger
\left(
\begin{array}{@{\,}cccc@{\,}}
\alpha&0&0&0\\
0&\beta&0&0\\
0&0&-\beta&0\\
0&0&0&-\alpha
\end{array}
\right) {\rm V}.
\end{equation}
This is an eigenvalue problem. The Hermitian matrix $A$ has the eigenvalues $\pm \alpha,\pm \beta$. Then we should solve the following system of equations for $i=0, 1, 2, 3$.
\begin{equation}
\label{last}
\alpha|v_{0i}|^2+\beta|v_{1i}|^2-\beta|v_{2i}|^2-\alpha|v_{3i}|^2=h_i, 
\end{equation}
where $h_0=-3/4$, otherwise $h_i=1/4$.

As far as we only want to show the possibility of the measurement of the conventional entanglement witness by only measuring the spin polarizations in a single run NMR experiment, it is enough to give an example and to avoid the unnecessary complexities due to a general proof.

Suppose that the state to be measured is a maximally entangled state. Measurement with the conventional witness observable $W$ gives $\langle W \rangle =-1/2$. We impose the observable $\tilde W$ to give the same value. We observe that for the maximally entangled state, ${\rm Z}_I$ and ${\rm Z}_S$ should take $\pm 1$. Recall that generally $c=1/4$. Then, as an example, we can choose $a_{\rm ex}=b_{\rm ex}=3/8$ and the corresponding unitary transformation is obtained as follows
\begin{equation}
 {\rm V}_{\rm ex}=\frac{1}{\sqrt 3}
\left(
\begin{array}{@{\,}cccc@{\,}}
0&1&e^{i\pi2/3}&e^{-i\pi2/3}\\
0&1&e^{-i\pi2/3}&e^{i\pi2/3}\\
0&1&1&1\\
{\sqrt 3}&0&0&0
\end{array}
\right).
\end{equation}
For different choices of ${zx}$, the elements of the matrix $A$ should be changed in a way that the unsimilar diagonal element $-3/4$ stands for the matrix component $A_{x+2z, x+2z}$. Then similar calculation shows that the components of ${\rm V}_{\rm ex}$ for different choices of ${zx}$ interchange according to each case.

Therefore, we showed that it is also possible to decompose the conventional entanglement witness to the separate but simultaneous measurement on the spin magnetizations. However, we emphasize that still the entanglement witness, which we introduced in this work is more handy as it covers all the cases for different $z$ and $x$, by just evaluation of the function $F$ with the absolute values without additional requirements on the different experimental operations. Whereas, as far as the conventional entanglement witness is concerned, different sets of $a$, $b$ and more over different unitary transformations are required for different choices of $z$ and $x$. In other words, by only changing the signs of the absolute values $a$ and $b$ but with a fixed unitary transformation, generally it is impossible to get the equality for the expectation values of $W$ and $\tilde W$. This can be proved by contradiction.

Suppose that it would be possible to have the equality with a fixed unitary transformation but just changing the signs of the absolute values. Then the equality (\ref{A}) should also be satisfied with a fixed ${\rm V}$ but different signs for $\alpha$ and $\beta$. Recall that the diagonal elements of $A$ have to interchange in accordance with the values of $z$ and $x$. All of them if applied to the system of equations (\ref{last}) result in some contradiction. Particularly two resultant equations from the system of equations for different $z$ and $x$ are as follows
\begin{equation}
\label{first}
\alpha|v_{00}|^2+\beta|v_{10}|^2-\beta|v_{20}|^2-\alpha|v_{30}|^2=-\frac{3}{4},
\end{equation}
\begin{equation}
\label{second}
\alpha|v_{00}|^2+\beta|v_{10}|^2-\beta|v_{20}|^2-\alpha|v_{30}|^2=-\frac{1}{4},
\end{equation}
where (\ref{first}) is for the case $a=|a|$ and $b=|b|$ but (\ref{second}) is for the case $a=-|a|$ and $b=-|b|$. The clear contradiction between (\ref{first}) and (\ref{second}) leads to the result that the original assumption can not be true.

We conclude as follows. The conventional entanglement witness is measurable in a single NMR experiment and by only measuring the spin magnetizations, provided that any required unitary transformation can be applied prior to the measurement. However, the unitary transformation has to be changed in accordance to the choices of $z$ and $x$ and in this sense the new entanglement witness, which we introduced in this contribution is proved to have its favorite advantage.

\end{document}